\documentclass[10pt,journal,compsoc]{IEEEtran}

\usepackage{amsmath,amsfonts}
\usepackage{algorithmic}
\usepackage{algorithm}
\usepackage{array}
\usepackage[caption=false,font=normalsize,labelfont=sf,textfont=sf]{subfig}
\usepackage{textcomp}
\usepackage{stfloats}
\usepackage{url}
\usepackage{verbatim}
\usepackage{graphicx}
\usepackage{cite}

\usepackage{microtype}

\usepackage{graphicx}
\usepackage{ragged2e}
\usepackage[justification=centering]{caption}

\usepackage{amsmath}
\usepackage{booktabs}
\usepackage{multirow}
\usepackage{array}
\usepackage{tabularx}

\usepackage{tikz}

\usetikzlibrary{decorations.pathmorphing}

\usepackage[edges]{forest}
\usepackage{xcolor}

\usepackage{amssymb}
\usepackage{bbding}
\usepackage{makecell}
\usepackage{capt-of}

\usepackage{url}
\usepackage{hyperref}
\hypersetup{
    colorlinks=true,
    linkcolor=blue,
    filecolor=magenta,      
    urlcolor=cyan,
}
\usepackage{hyperxmp}

\usepackage{diagbox}

\usepackage{pifont}
\usepackage{tcolorbox}
\newcommand{\summary}[1]{
    \begin{center}
    \begin{tcolorbox}[colback=gray!15, colframe=black, boxsep=-0.15cm, middle=-0.15cm]
        \textbf{\ding{46} Summary}
        $\blacktriangleright$
        {#1}
    $\blacktriangleleft$
    \end{tcolorbox}
    \end{center}
}

\newcommand{\finding}[1]{
    \begin{center}
    \begin{tcolorbox}[colback=white!15, colframe=black, boxsep=-0.15cm, middle=-0.15cm]
        \textbf{\ding{46} Finding}
        $\blacktriangleright$
        {#1}
    $\blacktriangleleft$
    \end{tcolorbox}
    \end{center}
}

\usepackage{longtable}
\usepackage{pdflscape}
\usepackage{makecell}
\usepackage{adjustbox}
\usepackage{caption}

\usepackage{graphicx}
\usepackage[normalem]{ulem}

\usepackage{color}

\newcommand{\ours}[1]{\textsc{SMNP}}

\def\BibTeX{{\rm B\kern-.05em{\sc i\kern-.025em b}\kern-.08em
    T\kern-.1667em\lower.7ex\hbox{E}\kern-.125emX}}

\begin{document}

\title{Understanding before Naming! Enhancing LLM-based Method Name Prediction with Code Summarization}

\author{Wei Liu, Weisong Sun, Tingting Xu, Hanwei Qian, Yi Zhao, Chunrong Fang, Xia Feng

\IEEEcompsocitemizethanks{

\IEEEcompsocthanksitem Wei Liu and Tingting Xu are with the Faculty of Data Science, City University of Macau, Macao, China. E-mail: \{D24092110297, D25092110136\}@cityu.edu.mo.

\IEEEcompsocthanksitem Xia Feng is with the School of Cyberspace Security (School of Cryptography), Hainan University, Haikou 570228, China. E-mail: xiafeng@hainanu.edu.cn

\IEEEcompsocthanksitem Weisong Sun is with the College of Computing and Data Science, Nanyang Technological University, Singapore 639798, Singapore.
E-mail: weisong.sun@ntu.edu.sg.

\IEEEcompsocthanksitem Hanwei Qian, Yi Zhao, and Chunrong Fang are with the State Key Laboratory for Novel Software Technology, Nanjing University, Nanjing 210093, China. E-mail: qianhanwei@smail.nju.edu.cn, zhaoyi@bjev.com.cn, and fangchunrong@nju.edu.cn.

}

\thanks{Manuscript received April xxx, 2026; revised August xxx, 2026.}
}

\markboth{Transaction on Software Engineering,~Vol.~xxx, No.~xxx, xxx~2026}
{Shell \MakeLowercase{\textit{et al.}}: Understanding before Naming! Enhancing LLM-based Method Name Prediction with Code Summarization}

\maketitle

\begin{abstract}

Method names are fundamental to software code quality, directly impacting code comprehensibility, maintainability, and the efficiency of developer collaboration. Crafting high-quality method names is a significant challenge, requiring precise abstraction of functionality, which can be particularly difficult for novice developers. Method Name Prediction (MNP), which automatically predicts a name for a given method code snippet, is becoming a significant research focus in Software Engineering (SE). Recently, MNP methods based on large language models (LLMs) have shown the potential to replace manual naming, but their practical application still faces two challenges. The first challenge is that the prevailing evaluation relies on metrics that primarily assess surface-level token overlap between predicted and reference names. This evaluation paradigm often fails to align with nuanced human judgment, limiting its practical usefulness. The second challenge is that current LLM-based MNP methods generate method names that deviate from the human-curated naming method. Instead, they operate as a prompt‑driven, probabilistic mapping that directly translates code representations and contextual information into method names, which is inconsistent with the ''understanding‑before‑naming'' idea in the human-curated naming method. The positive or negative impact of this inconsistency on the quality of the generated names is unknown.

To address the challenges discussed above, we conduct a series of empirical studies. Specifically, we first investigate the feasibility of using LLMs as evaluators to assess the quality of generated method names. We systematically compare 6 metric-based evaluators, 5 LLM-based evaluators, and 6 human evaluators. Experimental results show that, compared to the metric-based evaluators, the LLM-based evaluators, particularly those based on DeepSeek, are more consistent with the human evaluators, indicating that they can be used to assess the quality of method names. Furthermore, we investigate the impact of the ``understanding-before-naming'' idea on LLM-based MNP methods. We employ the validated LLM-based evaluator to compare the quality of method names generated by LLM-based MNP methods using two strategies: (1) the \textit{direct generation strategy}, which directly generates a method name from a given code snippet, and (2) the \textit{summarization-and-refinement strategy}, which first generates a code summary and then generates method names based on the summary, thereby simulating the human ''understanding‑before‑naming'' process.
The experimental results show that, overall, method names generated using the \textit{summarization-and-refinement strategy} are more semantically appropriate than those generated using the \textit{direct generation strategy}, exhibiting higher semantic coherence. Furthermore, we conduct an in-depth case study and identify three limitations in the naive implementation of the \textit{summarization-and-refinement strategy}: summarization semantic inaccuracy, refinement semantic misalignment, and semantic score proximity. We then propose a novel MNP approach, named \ours{}, that addresses these limitations through MNP-oriented summarization and chain-of-thought (CoT)–enhanced refinement. 
Large-scale experiments across 5 LLMs and 2 language datasets demonstrate the effectiveness of \ours{}, showing minimal sensitivity to LLM switching.

\end{abstract}

\begin{IEEEkeywords}
Method Name Prediction, Method Name Suggestion, Method Name Recommendation.
\end{IEEEkeywords}

\section{Introduction}
\label{sec:introduction}

\IEEEPARstart{I}{n} programming languages such as Java, Python, and C++, methods (also called functions) are essential building blocks that allow developers to encapsulate logic, promote code reuse, and enhance code organization and readability. Meaningful and succinct method names are important in software development, as they help developers quickly grasp the methods' key functionality~\cite{2014-deal-comprehension}. However, constructing high-quality method names is often challenging, especially for novice developers~\cite{2009-debug-name,2019-Code2seq,2019-code2vec,2016-Convolutional-Attention-Network}. Novice developers may struggle with the intricacies of code abstraction and with encapsulating a method's functionality in a concise, meaningful way. Method Name Prediction (MNP), also known as method name suggestion, is the task of suggesting appropriate names for methods based on the context of the method code~\cite{2019-Hierarchical-Attention-Networks,2022-Recommend-Method-Names-with-Global-Context,2021-DeepName,2020-Mnire}. By suggesting well-crafted and meaningful method names, MNP techniques guide best practices, helping developers learn effective naming conventions over time. The origins of MNP research can be traced back to the work of Allamanis et al.~\cite{2013-Mining-source-code-repositories-LM}. Their approach uses an n-gram language model, giga-token, trained on the GitHub Java corpus. With the rise of LLMs in SE, researchers increasingly explore LLM-inspired approaches for MNP. Fein et al.~\cite{2026-industrial-case-study} replace real method names with “dummy” tokens and prompt CodeT5, GPT-3.5-turbo, DeepSeek-Coder-v2, and Gemma-3 to directly generate candidate names from code contexts. UTGen~\cite{2025-UTGen} employs a structured prompt template with Code-Llama-7b-instruct to generate test method names, though its template is reused from multi-task scenarios.

Although recent LLM-based MNP methods have shown promise in automating manual naming, two challenges still hinder their practical deployment. The first challenge is that existing evaluation metrics, such as F1-score (F1) and Exact Match, primarily assess surface-level token overlap between predicted and reference names~\cite{2019-Code2seq,2019-code2vec,2019-RNN-two-improvement-strategies}. While these metrics are computationally convenient, they fail to capture semantic similarity or functional equivalence. This evaluation paradigm often misaligns with nuanced human judgment, creating a gap between reported model performance and developers' actual expectations in practice. For the second challenge, current LLMs generate method names in a manner inconsistent with the human-curated naming process. Specifically, they employ a prompt-driven mechanism that directly and probabilistically maps code representations and their surrounding context to method names, skipping the "understanding-before-naming" step that is fundamental to human naming~\cite{2026-industrial-case-study,2025-UTGen,2025-ContextCraft}. Humans, by contrast, first grasp the overall functionality of a code snippet and then devise a name that semantically reflects that functionality. The positive or negative impact of this inconsistency on naming quality remains unknown, motivating the need for a deeper understanding of how naming strategies influence generation quality.

For the first challenge, we investigate LLMs as evaluators to assess the quality of method names and validate their reliability. We systematically compare 6 metric-based evaluators, 5 LLM-based evaluators, and 6 human evaluators on the same model outputs. For automatic evaluation, we use widely used metrics including Precision, Recall, F1, ROUGE, BLEU, and Exact Match, computing their scores against reference names. For LLM-based evaluation, we instruct GPT-4o, DeepSeek, DeepSeek-Coder, Qwen, and Qwen-Coder to rate the quality of generated method names on a defined scale without providing reference names. For human evaluation, we invite 6 experienced experts and scholars to rate the semantic appropriateness of each generated name against the reference name. We then compute the Spearman correlation coefficient between the automatic metric scores and the human evaluation scores, and between the LLM-based evaluation scores and the human evaluation scores. The results show that LLM-based evaluations, particularly those driven by DeepSeek, correlate most strongly with human evaluations. These findings indicate that LLMs can serve as reliable evaluators of method name quality, offering a more human-aligned alternative to conventional metrics. For the second challenge, we investigate whether emulating the human "understanding-before-naming" process improves the quality of method names. We prompt the same five LLMs under two distinct conditions: (1) the \textit{direct generation strategy}, which directly generates a method name from a given code snippet, and (2) the \textit{summarization-and-refinement strategy}, which first generates a code summary and then generates a method name based on that summary, thereby simulating the human "understanding-before-naming" process. To evaluate the quality of the generated names, we apply the validated LLM-based evaluation method from the first challenge, using all five LLMs as evaluators. The results show that names generated under the \textit{summarization-and-refinement strategy} are more semantically appropriate than those under the \textit{direct generation strategy}, exhibiting greater semantic coherence and fewer naming deviations.

Using the LLM-based evaluation method validated in the first challenge, we analyze the method names generated in the second challenge and examine defects in both the generated summaries and the final method names. We further conduct an in-depth case study and identify three limitations in the naive implementation of the \textit{summarization-and-refinement strategy}: summarization semantic inaccuracy, refinement semantic misalignment, and semantic score proximity. To address these limitations, we propose a summary‑enhanced MNP (\ours{}) framework, which extends the \textit{summarization-and-refinement strategy} with two key innovations. First, we introduce a task‑oriented zero‑shot prompt for summary generation, which eliminates the need for task‑specific training examples while producing functionally accurate summaries. Second, we design a CoT prompting strategy that decomposes the naming process into three explicit reasoning steps: understanding the core purpose, identifying the key action, and determining the target entity. 
By making the intermediate reasoning explicit, \ours{} improves both prediction accuracy and interpretability. We evaluate \ours{} on widely used datasets in CSN-java and CSN-python. The experimental results show that \ours{} consistently outperforms the \textit{summarization-and-refinement strategy} across all five LLMs and both programming languages. 

In summary, we make the following contributions:

\begin{itemize}
\item We conduct a comprehensive study on using LLMs to evaluate method name quality. We find that LLM-based evaluators, particularly DeepSeek, align more closely with human judgments than metric-based evaluators.

\item We systematically examine the \textit{summarization-and-refinement strategy} for MNP. We find that it yields more semantically appropriate and coherent method names than \textit{direct generation strategy}, but it has inherent limitations.

\item Motivated by these findings, we propose a novel \ours{} framework, which addresses the limitations of the \textit{summarization-and-refinement strategy} and further improves the semantic quality of generated method names.

\item We release our dataset and code to facilitate reproducibility and enable future research in LLM-based method name evaluation and prediction~\url{https://github.com/software-theorem/SMNP}.
\end{itemize}

\section{Background and Related Work}
\label{sec:background}
\subsection{Method Name Prediction}
MNP can automatically generate appropriate method names from given code snippets, which is particularly beneficial for novice developers. In addition, it can be used to verify whether method names accurately reflect their implementations. In software maintenance, it is common for changes in code functionality to occur without corresponding updates to method names~\cite{2023-JIT-Method-Name-Updating}. Over the past decade, MNP has remained an important research topic in SE. The origins of MNP research can be traced back to the work of Allamanis et al.~\cite{2013-Mining-source-code-repositories-LM}. They propose an n-gram language model, giga-token, trained on the GitHub Java corpus. Their analysis reveals that method names are more predictable than type names and variable names, providing fundamental motivation for subsequent research. 

In recent years, the emergence of LLMs has opened new avenues for MNP research~\cite{2022-Multilingual-training-for-Software-Engineering,2023-AUMENA,2024-ijcnn-mn-framework}. 
Unlike previous approaches that require training from scratch, LLM-based methods adopt a prompt-driven approach, generating method names directly through prompt design without any task-specific training or parameter updates. Fein et al.~\cite{2026-industrial-case-study} employ models such as CodeT5, GPT-3.5-turbo, DeepSeek-Coder-v2, and Gemma-3 for method naming. Specifically, they replace the actual method name with a "dummy" token and feed the surrounding code context into the model, letting the LLM directly generate method names from the code. UTGen~\cite{2025-UTGen} leverages Meta's Code-Llama-7B-instruct model to generate unit test method names. It interacts with the LLM using a multi-task generic template (similar to a Chain-of-Thought prompt template), guiding the model to generate test method names based on the test body content. ContextCraft~\cite{2025-ContextCraft} retrieves semantically nearest historical examples from a corpus based on the input functional description, performs three types of context augmentation, and constructs a context-rich prompt, which is then fed into mainstream LLMs (e.g., ChatGPT-3.5, ChatGPT-4, ChatGPT-4o, Gemini-1.5, Llama-3) in a few-shot manner to generate method names. In this paper, we propose a novel MNP method, which first generates a code summary and then generates method names based on the summary, thereby simulating the human "understanding-before-naming" process. To address the limitations of this basic design, we then propose \ours{}, which enhances the approach through MNP-oriented summarization and chain-of-thought (CoT)-enhanced refinement.

\subsection{Code summarization}
Code summarization is the task of automatically generating natural-language summaries (also called comments) for code snippets~\cite{2023-Prompt-Learning-Framework-for-Code-Summarization,2023-few-shot-code-summarization}. These summaries serve various purposes, such as explaining code functionality and aiding comprehension~\cite{2022-practitioners-expectations, 2020-deep-code-comment, 2024-llm-few-shot-summarizers}. Research in this area dates back to 2010, when Haiduc et al.~\cite{2010-supporting-program-comprehension} first applied automated text summarization techniques to source code. Following the success of Neural Machine Translation (NMT) in natural language processing~\cite{2015-nmt-jointly-learning,2014-nmt-encoder-decoder-approaches}, many studies adopted the encoder–decoder architecture for code summarization~\cite{2020-transformer-summarization, 2020-comment-translation, 2020-retrieval-based-neural, 2023-extractive-and-abstractive, 2022-Evaluation-Neural-Code-Summarization, 2024-esale}. In recent years, research on LLM-based code summarization has grown rapidly~\cite{2025-code-summarization-era-of-llm}. Fried et al.~\cite{2023-inCoder} introduce InCoder and evaluate it in a zero-shot setting on the CodeXGLUE dataset; while it performed well, fine-tuned smaller models, such as CodeT5, still surpassed the zero-shot approach. To address concerns over code leakage when using commercial LLMs, Su et al.~\cite{2024-distilled-GPT} apply knowledge distillation to train smaller models from GPT-3.5, achieving comparable summarization quality. Ahmed et al.~\cite{2024-automatic-semantic-augmentation} further propose augmenting few-shot examples with semantic facts extracted from code. Sun et al.~\cite{2023-sun-arxiv-Summarization} design heuristic questions to collect GPT feedback and identify prompts that elicit in-distribution summaries. Rukmono et al.~\cite{2023-achieving-high-level-software-component} tackle LLMs' reasoning limitations by incorporating CoT prompting strategies. Sun et al.~\cite{2025-code-summarization-era-of-llm} conduct a comprehensive study on LLM-based code summarization. They find that GPT-4 evaluation aligns most closely with human judgment in code summarization, and that advanced prompting techniques do not necessarily outperform zero-shot prompting. In this paper, we build upon previous LLM-based code summarization methods, which focus on generating comments for developers rather than optimizing for method name generation. To address this, we leverage LLMs to generate code summaries and adapt them into a form suitable for method name prediction, thereby proposing MNP-oriented summarization.

\section{Study Design}
\label{sec:study_design}
\subsection{Research Questions}
To systematically investigate the two challenges identified above, this study aims to answer the following research questions (RQs). 

\textbf{RQ1: What is the viability of LLMs as evaluators for MNP?}

Conventional evaluation of MNP relies on metrics such as Precision, Recall, F1, ROUGE, Exact Match, and BLEU to measure the similarity between a predicted name and a single reference name~\cite{2019-code2vec,2019-Code2seq}. This paradigm assumes the reference name is an authoritative standard. However, empirical evidence shows that reference names in widely-used datasets can be ambiguous, overly generic, or semantically incomplete~\cite{2024-Exploring-mnp}. Consequently, techniques that produce a functionally correct but lexically divergent name may be unfairly penalized by these reference-dependent metrics, resulting in a distorted assessment of their true capabilities. This RQ systematically investigates the viability of employing LLMs as reference-free, semantic-aware evaluators for method name quality. We aim to achieve a more robust and reliable assessment, thereby overcoming the fundamental limitation of depending on potentially imperfect references.

\textbf{RQ2: What is the comparative effectiveness of the \textit{summarization-and-refinement strategy} versus the \textit{direct generation strategy} for LLM-based MNP?}

This RQ is examined by comparing our proposed \textit{summarization-and-refinement strategy} with the conventional \textit{direct generation strategy}. The former first generates a natural language summary of the code and then derives a method name, whereas the latter maps source code directly to a method name. This comparison directly tests the impact of this inconsistency on generation quality, specifically whether emulating the human "understanding-before-naming" process yields positive or negative effects. Accordingly, we aim to comprehensively evaluate, both quantitatively and qualitatively, whether the summarization-and-refinement strategy improves the accuracy, relevance, and semantic consistency of the LLM-generated method names.

\subsection{Experimental LLMs}
We select five LLMs as representatives in our experiments, including open-source and proprietary models, as well as general-purpose and code-specific models.

\textbf{GPT.} We employ GPT-4o-mini from the GPT-4o family by OpenAI as a leading closed-source general-purpose LLM~\cite{2022-ChatGPT}. Pre-trained on large-scale text and code corpora, it delivers strong natural language and code understanding and generation. Although its exact parameter count is unknown, its broad adoption provides a solid baseline for performance in this study.

\textbf{DeepSeek.} For open-source models specializing in complex reasoning, we adopt DeepSeek-Reasoner from DeepSeek AI~\cite{2025-deepseek-r1}. Trained with fine-tuning and reinforcement learning, the model improves multi-step logical deduction and analytical capabilities. The model's strength in complex reasoning enables it to capture nuanced semantic information from code.

\textbf{DeepSeek‑Coder.} From the DeepSeek‑Coder series, we use deepseek‑coder‑6.7b‑instruct~\cite{2024-DeepSeek-Coder}. DeepSeek-Coder is a 6.7B-parameter, instruction-tuned model trained from scratch on 2 trillion tokens (87\% code). With its strong capability in code generation and comprehension, it serves as a representative open-source code expert in our evaluation.

\textbf{Qwen.} We select Qwen2.5-7B-Instruct~\cite{2025-Qwen} from the Qwen family. This 7-billion-parameter model is trained on diverse corpora, balancing natural language understanding, reasoning, and instruction-following. Its inclusion broadens the diversity of model architectures in our evaluation, allowing us to examine whether our approach remains effective across models with varying training objectives and capabilities.

\textbf{Qwen‑Coder.} We select Qwen2.5-Coder-7B-Instruct~\cite{2025-Qwen-Coder} from the Qwen-Coder family. This model is specifically designed for code-related tasks and is trained on large-scale code data, including source code, text-code grounding, and synthetic data. As a result, it achieves strong capabilities in code generation, code reasoning, and code fixing. 

\textbf{Model Settings.} We set the temperature to 0 across all experiments to ensure deterministic outputs from the LLMs. This controlled setup allows us to isolate and examine the effects of key variables: evaluation methods, prompting techniques, and programming language types.

\subsection{Experimental Datasets}

\textbf{CodeSearchNet (CSN).}
To evaluate the performance of LLMs in generating names for target method code across different languages, we utilize the multilingual dataset CSN corpus~\cite{2021-CodeXGLUE}, which has been employed in existing MNP studies~\cite{2022-HPG}. The corpus comprises approximately 6 million functions from six programming languages, including Go, Java, JavaScript, PHP, Python, and Ruby. In this study, we use Java and Python samples as experimental datasets because they are more commonly used in MNP tasks and cover two mainstream naming conventions (camelCase and snake\_case). We utilize the clean version of the CSN corpus, as provided by Lu et al.~\cite{2021-CodeXGLUE}, in CodeXGLUE. We follow ~\cite{2025-code-summarization-era-of-llm} and randomly select 200 samples for each programming language from the test set of this corpus for experiments.

\section{Empirical Study Results and Findings}
\label{sec:empirical_study_results_and_findings}

\subsection{RQ1: What is the viability of LLMs as evaluators for MNP?}
\label{subsec:RQ1_LLMs_evaluators}

\subsubsection{Experimental Setup}

\textbf{1) Metric-based Evaluation.}
Existing MNP studies use various performance metrics to evaluate the effectiveness of an MNP technique. In performance evaluation, most studies primarily measure precision, recall, and F1~\cite{2023-AUMENA,2021-code-transformer}, while some use exact-matching accuracy~\cite{2022-Neural-Model,2019-RNN-two-improvement-strategies}.

\noindent\textbf{$\bullet$ Precision.} Precision refers to the proportion of correctly predicted method name tokens from the total number of predicted method name tokens. It measures the accuracy of the MNP technique.

\noindent\textbf{$\bullet$ Recall.} Recall is the proportion of correctly predicted method name tokens to the total number of tokens in the target method name. It measures the recall ability of the MNP technique.

\noindent\textbf{$\bullet$ F1.} The F1 is the harmonic mean of precision and recall and provides a comprehensive measure of system performance. A high F1 indicates a good balance between precision and recall.

In the evaluation process, the reference/ground-truth method name ($N_{g}$) and the predicted method name ($N_{p}$) are treated as a pair, and the following calculations are performed:
\begin{equation}
   P = Precision = \frac{|token(N_{g}) \cap token(N_{p})|}{|token(N_{p})|} 
    \label{equ:Precision}
\end{equation}

\begin{equation}
     R = Recall = \frac{| token(N_{g}) \cap token(N_{p}) |}{| token(N_{g})|}
    \label{equ:Recall}
\end{equation}

\begin{equation}
    F1 = F1-score = 2 \times \frac{P \times R}{P + R}
    \label{equ:F1-score}
\end{equation}

where $token(\cdot)$ is a function that returns the (sub)tokens in the method name $\cdot$, and $|\cdot|$ denotes the number of tokens. Similar to F1, the Modified-F1 (F1**) score is also used to evaluate method names. The Modified-F1 (F1**) score is calculated from the modified unigram precision and unigram recall. This prevents models that repeatedly output subwords in the gold method name from receiving unreasonably high scores.

\textbf{$\bullet$ Exact Match(EM).} EM measures whether the predicted method name exactly matches the ground-truth method name. Specifically, if the predicted method name $N_p$ exactly matches the ground-truth method name $N_g$, the EM score is 1; otherwise, it is 0.

\begin{equation}
    EM = \begin{cases}
    1, & \text{if } N_{g} = N_{p} \\
    0, & \text{otherwise}
\end{cases}
\end{equation}

\textbf{$\bullet$ ROUGE.} ROUGE is a set of metrics originally designed for evaluating text summarization systems. In this work, it is used to measure the lexical overlap between predicted and ground-truth method names. Typical variants include ROUGE-N and ROUGE-L. ROUGE-N measures the n-gram overlap between the generated text and the reference text, where $n$ denotes the length of the n-gram. ROUGE-L evaluates their similarity based on the longest common subsequence (LCS).
\begin{equation}
    ROUGE-N = \frac{| token_{n}(N_{g}) \cap token_{n}(N_{p}) |}{|token_{n}(N_{g})|}  
\end{equation}
where $n$ denotes the length of the n-gram, $token_n(\cdot)$ denotes the set of n-grams extracted from the input text, $N_g$ denotes the ground-truth method name, and $N_p$ denotes the predicted method name.

\textbf{ROUGE-L.} ROUGE-L evaluates the similarity between two method names by considering the longest common subsequence (LCS), which captures the sequential order of tokens.~\cite{2019-sequence-GNN,2021-Keywords-Guided-MNG}. Its calculation formula is as follows:

\begin{equation}
    R_{lcs}=\frac{LCS(N_{g}, N_{p})}{|token(N_{g})|}
\end{equation} 
\begin{equation}
    P_{lcs}=\frac{LCS(N_{g}, N_{p})}{|token(N_{p})|} 
\end{equation}
\begin{equation}
   F_{lcs}=\frac{(1+\beta^{2})R_{lcs}P_{lcs}}{R_{lcs}+\beta^{2}P_{lcs}}  
\end{equation}
where $LCS(N_{g}, N_{p})$ is the length of the LCS of $N_{g}$ and $N_{p}$. $R_{lcs}$ and $P_{lcs}$ denote recall and precision. The last $F_{lcs}$ is used as the ROUGE-L score. $\beta$ is a parameter that adjusts the relative importance of precision and recall. If $\beta$ is set to  {1}, the $F_{lcs}$ represents the harmonic mean of precision and recall, and its calculation is the same as that of the F1.

\textbf{$\bullet$ BLEU.} BLEU is a commonly used metric in machine translation tasks to measure the similarity between candidate sentences and reference sentences. It evaluates the quality of a candidate sequence based on modified n-gram precision and a brevity penalty. In the MNP task, the predicted method name and the ground-truth method name are treated as the candidate sequence and the reference sequence, respectively. The BLEU score is calculated as follows:

\begin{equation}
    BLEU = BP \cdot \exp\left( \sum_{n=1}^{N} w_n \log p_n \right),
\end{equation}
where $p_n$ denotes the modified n-gram precision of order $n$, which is computed based on the n-gram overlap between the predicted method name $N_p$ and the ground-truth method name $N_g$. $w_n$ is the weight assigned to the n-gram precision of order $n$. The modified n-gram precision is computed by dividing the clipped count of matched n-grams by the total number of n-grams in the candidate sequence. To reduce the bias toward short candidate sequences, a brevity penalty is introduced:
\begin{equation}
    BP =
    \begin{cases}
        1, & \text{if } c > r, \\
        e^{(1-r/c)}, & \text{if } c \leq r,
    \end{cases}
\end{equation}
where $c=|token(N_p)|$ is the length of the predicted method name, and $r=|token(N_g)|$ is the length of the ground-truth method name. In this work, BLEU-4 is used, which considers 1-gram, 2-gram, 3-gram, and 4-gram matching between the predicted and ground-truth method names.

\textbf{2) Human Evaluation.}
We conduct human evaluations as the reference standard for assessing both metric-based and LLM-based evaluation results. To answer RQ1, we compare the correlations of both metric-based evaluation results and LLM-based evaluation results with human evaluations, aiming to determine whether LLMs can serve as reliable automated evaluators for MNP. We invite six volunteers with more than 5 years of software development experience and strong English proficiency to perform the evaluation. For each sample, we provide volunteers with the code snippet, the reference method name, and the method names generated by the LLMs. The reference method names and the LLM-generated method names are mixed and presented in random order. 

We follow the approach of previous studies~\cite{2025-Summarization-LLM, 2022-Evaluation-Neural-Code-Summarization} and ask volunteers to evaluate the quality of each method name based on two criteria (C1 and C2). The final score for each method name is the average of the six volunteers' scores.

\textbf{C1. Accuracy Score (1–5):} A good method name should satisfy three conditions: (1) clearly convey the primary purpose of the method and accurately reflect its core functionality; (2) be concise yet descriptive, avoiding unnecessary words or redundancy; (3) avoid using generic or ambiguous terms and ensure clarity and readability. The rating scale ranges from 1 to 5, where a score of 1 indicates that the method name is completely irrelevant, misleading, or does not reflect the method's functionality at all; scores of 2, 3, and 4 represent intermediate levels of quality; and a score of 5 indicates that the method name is perfectly aligned with the functionality, follows naming conventions, is concise yet fully descriptive, and avoids any ambiguity or redundancy.

\textbf{C2. Naming Format Check (1 or 0):} A score of 1 indicates that the method name follows the standard naming conventions of the target programming language (e.g., camelCase for Java, snake\_case for Python), while 0 indicates non-compliance.

\textbf{3) LLM-based Evaluation.} Inspired by recent works~\cite{2025-Summarization-LLM, 2024-Disinformation-Capabilities, 2023-nlg-evaluator}, we further investigate the feasibility of employing LLMs as evaluators for MNP. This evaluation strategy does not depend solely on the quality of reference method names and follows the same evaluation protocol and prompt as the human evaluation. For each sample, the code snippet, the reference method name, and the LLM-generated method names are provided to the LLM evaluators, which are then asked to rate each method name on a scale from 1 to 5, with higher scores indicating better naming quality.

\textbf{4) Datasets and Prompting Techniques.}
\label{sec:datasets_prompting_techniques}
We describe our dataset sampling and prompting technique as follows.

\textbf{Datasets.} In this RQ, to reduce the workload of human evaluation volunteers, we randomly select 50 samples from the Java and Python datasets, respectively, which means 100 samples in total.

\textbf{Prompting Techniques.} 
We employ zero-shot prompting to adapt the five LLMs to generate method names for code snippets. 
The prompt is designed to produce naming without intermediate semantic transformation. It directs the model to act as an experienced developer and generate a method name directly from the provided source code. Furthermore, the prompt articulates four key naming principles, each justified as follows:

\begin{enumerate}
    \item \textbf{Accurately reflecting the core functionality:} This principle directs the model to distill the code’s essential purpose into the name, thereby enabling the generated identifier to serve as a clear semantic summary of the method’s behavior.

    \item \textbf{Adhering to language-specific conventions (e.g., camelCase for Java, snake\_case for Python):} This principle ensures the model follows the syntactic norms of the target programming language, promoting consistency with existing codebases and enhancing developer familiarity.

    \item \textbf{Balancing conciseness with descriptiveness:} This principle guides the model to optimize name length against information density, avoiding excessive verbosity while preserving sufficient detail for comprehension.

   \item \textbf{Avoiding vague or generic terms:} This principle encourages the model to produce distinctive and semantically precise names, reducing ambiguity and improving the name’s utility in downstream development tasks.
\end{enumerate}

The prompt embeds four principles that systematically guide the model from distinct yet complementary dimensions, respectively addressing semantic accuracy, syntactic convention, expressive efficiency, and lexical specificity. By jointly enforcing these constraints, the prompt optimizes the model’s output to align with human naming conventions. This ensures that generated method names are not only functionally descriptive and syntactically correct but also concise, unambiguous, and developer-friendly.

\subsubsection{Experimental Results}

\begin{table}[!t]
    \centering
    \small
    \caption{Human evaluation scores for reference and LLM-generated method names. The value in parentheses represents the percentage increase or decrease relative to the score of the corresponding reference.}
    \label{tab:human_eval}
    \begin{tabular}{lcc}
    \hline
    \multicolumn{1}{l}{\multirow{2}{*}{Name from}} & \multicolumn{2}{c}{Human Evaluation Score} \\ \cline{2-3}
    \multicolumn{1}{c}{} & Java & Python \\ \hline
    Reference  & 3.53 &  3.25  \\ \hline
    GPT-4o & 4.37(+23.89\%) & 4.13(+26.97\%) \\ 
    DeepSeek & 4.20(+18.98\%) & 4.03(+24.10\%) \\ 
    DeepSeek-Coder & 4.07(+15.30\%) & 3.80(+16.92\%) \\ 
    Qwen & 3.70(+4.91\%) & 3.70(+13.74\%) \\ 
    Qwen-Coder & 4.18(+18.32\%) & 3.82(+17.44\%) \\ \hline
    \end{tabular}
\end{table}

\textbf{1) Human Evaluation Results. }Table~\ref{tab:human_eval} shows the human evaluation scores for reference method names and method names generated by five LLMs. Observe that the scores of reference method names in Java and Python are 3.53 and 3.25, respectively, suggesting that the quality of the reference method names is limited. Therefore, given their limited quality, evaluation approaches that rely primarily on reference comparisons may not reliably reflect the quality of LLM-generated method names. 

Among the five LLMs, GPT-4o achieves the highest scores in both Java (4.37) and Python (4.13), demonstrating the strongest capability in method name generation. All LLMs outperform the reference names, underscoring their effectiveness in code semantic understanding.

\finding{According to human evaluation, the quality of reference method names in existing datasets is limited. Method names generated by all five LLMs significantly surpass the reference names in quality, with GPT-4o demonstrating the most substantial improvement.}

 \begin{table*}[ht]
    \centering
    \tiny
    \caption{Metric-based and LLM-based evaluation scores for reference and LLM-generated method names.}
    \label{tab:automated_evaluation_scores}
    \resizebox{\textwidth}{!}{
    \begin{tabular}{llcccccccccccc}
    \toprule
    \multirow{2}{*}{\textbf{Language}} & \multirow{2}{*}{\textbf{Name from}} 
    & \multicolumn{6}{c}{\textbf{Metric-based Evaluation}} 
    & \multicolumn{5}{c}{\textbf{LLM-based Evaluation}} 
    & \multirow{2}{*}{\textbf{Human}} \\
    \cmidrule(lr){3-8} \cmidrule(lr){9-13}
    & & Precision & Recall & F1 & BLEU & ROUGE & Exact Match & ChatGPT & DeepSeek & DeepSeek-Coder & Qwen & Qwen-Coder \\
    \midrule
    \multirow{6}{*}{Java}		
    
    & Reference             & $\backslash$ & $\backslash$ & $\backslash$ & $\backslash$ & $\backslash$ & $\backslash$ & 3.04 & 2.84 & 2.99 & 2.7 & 2.5 & 3.53 \\
    & GPT-4o     & 33.87 & 58.23 & 40.59 & 10.77 & 39.28 & 2.00 & \textbf{4.64} & \textbf{4.28} & \textbf{4.03} & \textbf{4.12} & \textbf{4.08} & \textbf{4.37} \\ 
    & DeepSeek      & \textbf{48.77} & \textbf{62.57} & \textbf{51.81} & \textbf{15.95} & \textbf{51.67} & \textbf{22.00} & 3.90 & 4.04 & 3.69 & 3.44 & 3.28 & 4.20 \\
    & DeepSeek-Coder& 39.87 & 58.23 & 44.61 & 12.53 & 44.19 & 6.00 & 3.82 & 3.42 & 3.75 & 3.44 & 3.32 & 4.07 \\
    & Qwen & 34.97 & 47.33 & 38.07 & 12.12 & 37.48 & 2.00 & 3.50 & 2.72 & 3.57 & 3.12 & 3.06 & 3.70 \\
    & Qwen-Coder & 33.67 & 52.40 & 38.89 & 11.09 & 37.00 & 4.00 & 4.06 & 3.68 & 3.89 & 3.64 & 3.52 & 4.18 \\
     							      					
    \midrule
    \multirow{6}{*}{Python}
    & Reference             & $\backslash$ & $\backslash$ & $\backslash$ & $\backslash$ & $\backslash$ & $\backslash$ & 2.74 & 2.64 & 2.84 & 2.42 & 2.52 & 3.25 \\
    & GPT-4o     & 22.67 & 41.67 & 28.04 & 6.85 & 28.04 & 0.00 & \textbf{4.50} & 3.80 & \textbf{4.13} & \textbf{3.76} & \textbf{4.00} & \textbf{4.13} \\
    & DeepSeek      & \textbf{38.63} & \textbf{56.50} & \textbf{43.83} & \textbf{11.79} & \textbf{43.83} & \textbf{12.00} & 3.84 & \textbf{4.10} & 3.65 & 3.22 & 3.54 & 4.03 \\
    & DeepSeek-Coder& 27.97 & 41.17 & 31.94 & 9.23 & 31.89 & 2.00 & 3.78 & 3.08 & 3.91 & 3.12 & 3.38 & 3.80 \\
    & Qwen          & 31.47 & 42.33 & 34.22 & 9.34 & 33.55 & 4.00 & 3.54 & 2.92 & 3.67 & 3.22 & 3.26 & 3.70 \\
    & Qwen-Coder    & 27.40 & 46.00 & 32.65 & 8.12 & 32.65 & 0.00 & 3.68 & 3.12 & 3.92 & 3.32 & 3.44 & 3.82 \\
    						
    \bottomrule
    \end{tabular}
    } 
\end{table*}

\textbf{2) Automated Evaluation Results. }Table~\ref{tab:automated_evaluation_scores} presents the comprehensive results for the method name generation, encompassing metric-based evaluation, LLM-based evaluation, and human evaluation. For both Java and Python, DeepSeek achieves the highest scores in metric-based evaluation, yielding the best Precision, Recall, F1, BLEU, ROUGE, and Exact Match values. This indicates that method names generated by DeepSeek are more lexically similar to the reference names, with greater token overlap and closer alignment with local lexical patterns. In contrast, GPT-4o attains the highest scores in LLM-based evaluation and human evaluation. This finding suggests that GPT-4o's generated method names are preferred in terms of semantic adequacy, readability, and naming naturalness, even when they do not exactly match the reference names. Metric-based evaluation primarily measures lexical similarity between generated method names and the reference names, rather than assessing the overall quality of the generated names. ROUGE, Precision, Recall, and F1 reward token overlap, whereas BLEU is more sensitive to exact local n-gram matches and their sequence order. Exact Match imposes an even stricter criterion, requiring the generated name to be identical to the reference. Consequently, these metrics may favor names that are lexically close to the reference and may underestimate semantically correct, natural, or more readable names that use different expressions.
\finding{According to the evaluation results, DeepSeek achieves the highest scores in metric-based evaluation. In contrast, GPT-4o attains the highest scores in LLM-based evaluation and human evaluation. Moreover, metric-based evaluation mainly captures lexical similarity to reference names, whereas LLM-based and human evaluations are more effective at capturing semantic correctness and overall naming quality.}

\textbf{3) Correlation between Automated Evaluation and Human Evaluation. } 
Based on these findings, we speculate that LLM-based evaluation may be more suitable for assessing the quality of LLM-generated method names than metric-based evaluation. To test this speculation and provide statistical evidence for these observations, following prior research \cite{2022-Evaluation-Neural-Code-Summarization,2021-metric-reassessing}, we calculate Spearman's rank correlation coefficient $\rho$ between metric-based evaluation methods and LLM-based evaluation methods against human evaluation results. The Spearman’s correlation coefficient $\rho \in [ -1, 1] $ is commonly used to measure the monotonic relationship between two rankings, where an absolute value closer to 1 indicates a stronger correlation~\cite{1999-Spearman-correlation-coefficient}. Specifically, $\rho > 0$ signifies a positive correlation, $\rho < 0$ a negative correlation, and $\rho = 0$ suggests no significant monotonic relationship~\cite{2007-statistics-without-maths-psychology}. The $p$-value is used to determine whether the observed correlation is statistically significant. By comparing the $p$-value to a preset significance level (typically 0.05), we can determine whether to reject the null hypothesis and decide whether the correlation is statistically significant. As shown in Table~\ref{tab:correlation_coefficient}, the correlation results reveal two key observations. Compared with metric-based evaluation, LLM-based evaluation demonstrates more consistent agreement with human evaluation, suggesting that LLM evaluators are better able to assess the semantic adequacy and naturalness of generated method names. Additionally, among LLM-based evaluation methods, DeepSeek exhibits the strongest and most consistent correlation with human evaluation, indicating that DeepSeek-based judgments align most closely with human preferences.

\begin{table*}[ht]
\centering
\tiny
\caption{Spearman’s correlation coefficient $\rho$ with the p-value (values in parentheses) between the results of each evaluation method and human evaluation.}
\label{tab:correlation_coefficient}
\resizebox{\textwidth}{!}{
\begin{tabular}{llccccccccccc}
\toprule
\multirow{2}{*}{\textbf{Language}} & \multirow{2}{*}{\textbf{Name from}} 
& \multicolumn{6}{c}{\textbf{Metric-based Evaluation}} 
& \multicolumn{5}{c}{\textbf{LLM-based Evaluation}}  \\
\cmidrule(lr){3-8} \cmidrule(lr){9-13}
& & Precision & Recall & F1 & BLEU & ROUGE  & Exact Match & GPT-4o & DeepSeek & DeepSeek-Coder & Qwen & Qwen-Coder \\
\midrule
\multirow{6}{*}{Java}		

& Reference              & $\backslash$ &$\backslash$ &  $\backslash$ &  $\backslash$ &  $\backslash$ &  $\backslash$ & \textbf{0.83(.00)} & 0.74(.00)& 0.67(.00) & 0.60(.00) & 0.67(.00) \\
& GPT-4o        & -0.10(.49) & 0.21(.15) & 0.02(.90) & -0.10(.50) & 0.04(.76) & -0.06(.70) & 0.57(.00) & \textbf{0.60(.00)} & 0.21(.15) & 0.38(.01) & 0.46(.00) \\
& DeepSeek         & -0.10(.50) & 0.13(.37) & -0.03(.86) & 0.00(.98) & -0.03(.83)  & -0.03(.82) & 0.79(.00) & 0.74(.00) & 0.53(.00) & 0.69(.00) & \textbf{0.80(.00)} \\
& DeepSeek-Coder   & 0.10(.48) & 0.18(.20) & 0.15(.28) & 0.17(.23) & 0.12(.40) & 0.25(.08) & 0.78(.00) & \textbf{0.90(.00)} & 0.39(.00) & 0.60(.00) & 0.71(.00) \\
& Qwen             & 0.19(.18) & 0.45(.00) & 0.30(.03) & 0.29(.04) & 0.30(.04) & 0.19(.18) & 0.78(.00) & \textbf{0.89(.00)} & 0.43(.00) & 0.67(.00) & 0.71(.00) \\
& Qwen-Coder       & 0.34(.02) &\textbf{ 0.48(.00)} & 0.40(.00) & 0.29(.04) & 0.43(.00) & 0.25(.08) & 0.70(.00) & \textbf{0.81(.00)} & 0.28(.05) & 0.60(.00) & 0.72(.00) \\
 							      					
\midrule
\multirow{6}{*}{Python}
& Reference             & $\backslash$ &$\backslash$ & $\backslash$ & $\backslash$ & $\backslash$ & $\backslash$ & 0.71(.00) & \textbf{0.84(.00)} & 0.40(.00) & 0.51(.00) & 0.52(.00) \\
& GPT-4o       & -0.02(.91) & 0.03(.85) & 0.01(.94) & 0.00(.99) & 0.01(.94) &  $\backslash$ & 0.30(.04) & 0.53(.00) & 0.09(.52) & 0.32(.02) & \textbf{0.58(.00)} \\
& DeepSeek        & 0.10(.48) & 0.19(.18) & 0.16(.26) & 0.13(.36) & 0.16(.26) & 0.04(.77) & 0.59(.00) & \textbf{0.62(.00)} & 0.45(.00) & 0.49(.00) & 0.57(.00) \\
& DeepSeek-Coder  & 0.19(.20) & 0.23(.11) & 0.21(.15) & 0.23(.10) & 0.21(.15) & 0.02(.86) & 0.69(.00) & \textbf{0.70(.00)} & 0.29(.04) & 0.56(.00) & 0.64(.00) \\
& Qwen            & 0.18(.22) & 0.23(.11) & 0.20(.16) & 0.18(.22) &  0.21(.14) & 0.12(.42) & 0.54(.00) & \textbf{0.81(.00)} & 0.41(.00) & 0.62(.00) & 0.68(.00) \\
& Qwen-Coder      & 0.18(.20) & 0.39(.00) & 0.31(.03) & 0.21(.15) &0.31(.03)  &  $\backslash$ & 0.63(.00) & \textbf{0.69(.00)} & 0.37(.01) & 0.51(.00) & 0.60(.00) \\
 						
\bottomrule
\end{tabular}
}
\end{table*}

\summary{LLM-based evaluation correlates more consistently with human evaluation than metric-based evaluation. Among LLM-based methods, DeepSeek shows the strongest correlation with human evaluation.}

\subsection{RQ2: What is the comparative effectiveness of the \textit{summarization-and-refinement strategy} versus the \textit{direct generation strategy} for LLM-based MNP?}
\label{subsec:RQ2_comparative_effectiveness}
\subsubsection{Experimental Setup.}

Building on the aforementioned experiments, which have validated the feasibility of using LLMs as evaluators for method names, we further investigate whether incorporating \textit{summarization-and-refinement strategy} can improve the quality of LLM-generated method names. The experiment continues with the previously randomly selected 200 samples, employing a zero-shot prompting approach, and designs and compares the following two generation strategies.

\begin{figure}[!t]
    \centering
    \includegraphics[width=\columnwidth]{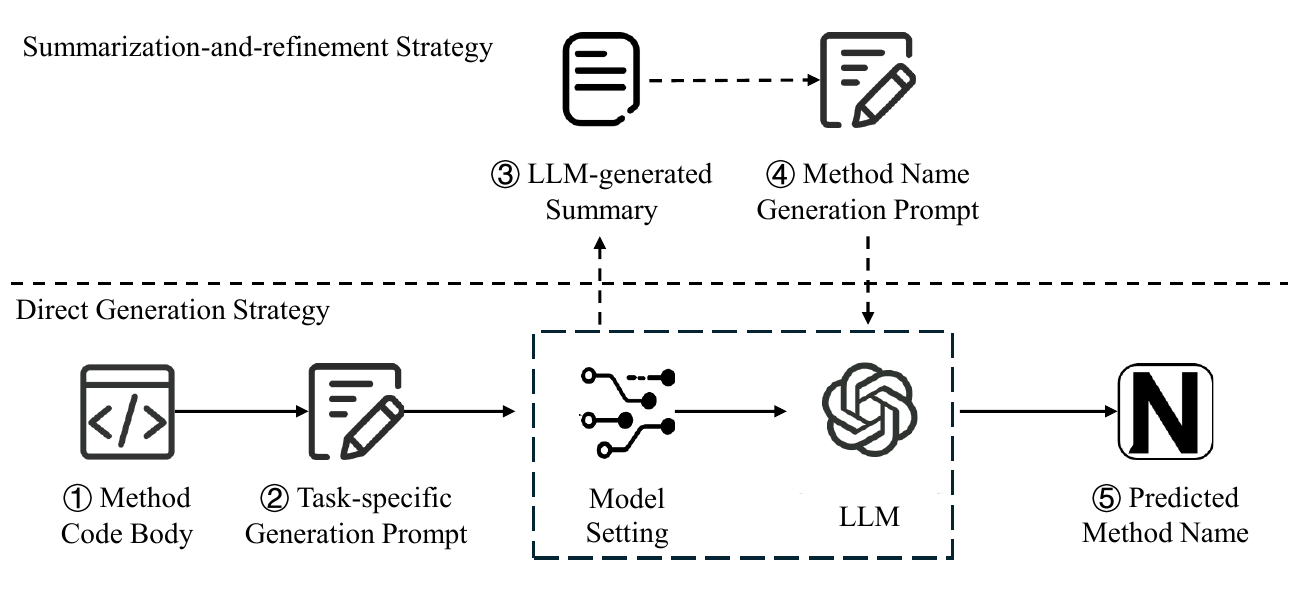}
    \caption{\protect\raggedright Workflow of LLM-based MNP under two generation strategies: \textit{direct generation strategy} and \textit{summarization-and-\allowbreak refinement strategy}. Solid arrows indicate the main generation flow, while dashed arrows indicate the additional refinement flow used in the \textit{summarization-and-\allowbreak refinement strategy}.}
    \label{fig:workflow_llm_based_mnp_two_generation_strategies}
\end{figure}

\textbf{\textit{Direct Generation Strategy}:}
In the \textit{direct generation strategy}, the method code body is directly provided to the LLM together with the task-specific generation prompt described in the prompting techniques discussion in Section~\ref{sec:datasets_prompting_techniques}. This prompt follows a zero-shot setting and instructs the LLM to act as an experienced developer, infer the functionality and intent of the method from its implementation, and generate a concise and semantically appropriate method name. As shown by the solid-arrow path in Figure~\ref{fig:workflow_llm_based_mnp_two_generation_strategies}, this strategy corresponds to steps \textcircled{1}, \textcircled{2}, and \textcircled{5}, producing the predicted method name without introducing intermediate semantic representations or additional refinement steps.

\textit{\textbf{Summarization-and-refinement\ Strategy}}:
As shown in Figure~\ref{fig:workflow_llm_based_mnp_two_generation_strategies}, the \textit{summarization-and-refinement strategy} is designed to mimic the human process of naming method code bodies, which typically involves first understanding the functional purpose of the code and then formulating an appropriate method name based on that understanding. As indicated by the dashed-arrow path, this strategy corresponds to steps \textcircled{1}--\textcircled{5} and introduces an additional refinement flow beyond the \textit{direct generation strategy}. By explicitly separating code comprehension from name formulation, this strategy provides a clearer mapping from source code to method names and better aligns the generation process with human naming practices.

\textbf{(1) Code Summarization:}
This stage corresponds to steps \textcircled{1}--\textcircled{3} in Figure~\ref{fig:workflow_llm_based_mnp_two_generation_strategies}. In this stage, the task-specific generation prompt in step \textcircled{2} is instantiated as a code summarization prompt. Following Sun et al.~\cite{2025-Summarization-LLM}, we adopt a zero-shot prompt for this summarization task: \textit{``Please generate a short comment in one sentence for the following function:$<source\ code>$.''} Given the method code body and this prompt, the LLM is instructed to generate a functional summary of the code. The prompt guides the model to focus on the functional purpose of the method while omitting low-level implementation details, thereby producing a concise description of the method's intended behavior.

\textbf{(2) Method Name Generation:}
This stage corresponds to steps \textcircled{4}--\textcircled{5} in Figure~\ref{fig:workflow_llm_based_mnp_two_generation_strategies}. The input consists of the code summary generated in the previous stage (i.e., step \textcircled{3}) and a dedicated method name generation prompt. The prompt follows the same four naming principles as those used in the \textit{direct generation strategy}, but replaces the original method code body with the generated summary as input. Accordingly, the LLM generates the final method name solely based on the functional understanding encapsulated in the summary. By relying on the intermediate summary rather than the original code body, this stage encourages the model to reason and select vocabulary at the level of abstract functional description. This design allows us to examine whether introducing an explicit semantic summary can improve the quality of generated method names.
\begin{table*}[!t]
    \centering
    \tiny
    \tabcolsep=12pt
    \caption{Comparison of the \textit{direct generation strategy} (DGS) and the \textit{summarization-and-refinement strategy} (SRS) using the LLMs as evaluators}
    \label{tab:evaluation_direct_generation_strategy_vs_summarization-and-refinement_strategy}
    \begin{tabular}{llcccccc}
        \toprule
        \multirow{2}{*}{\textbf{Language}} 
        & \multirow{2}{*}{\textbf{Name Source}} 
        & \multicolumn{5}{c}{\textbf{LLM-based Evaluation}} \\
        \cmidrule(lr){3-8}
        & & GPT-4o & DeepSeek & DeepSeek-Coder & Qwen & Qwen-Coder & Average \\
        \midrule
        \multirow{11}{*}{Java}
        & Reference              & 2.94 & 2.67 & 3.07 & 2.57 & 2.49 & 2.75\\
        
        & GPT-4o\_DGS         & \textbf{4.49} & \textbf{4.04} & 3.78 & \textbf{3.93} & \textbf{3.87} & \textbf{4.02} \\
        & GPT-4o\_SRS         & 4.37 & 3.89 & \textbf{3.90} & 3.76 & 3.78 & 3.94 \\
        
        & DeepSeek\_DGS          & 3.88 & 3.98 & 3.54 & 3.43 & 3.25 & 3.62 \\
        & DeepSeek\_SRS          & \textbf{3.91} & \textbf{4.08} & \textbf{3.77} & \textbf{3.62} & \textbf{3.40} & \textbf{3.76} \\
        
        & DeepSeek-Coder\_DGS    & 3.73 & 3.24 & 3.62 & 3.28 & 3.26 & 3.43 \\
        & DeepSeek-Coder\_SRS    & \textbf{3.88} & \textbf{3.37} & \textbf{3.81} & \textbf{3.52} & \textbf{3.49} & \textbf{3.61} \\
        
        & Qwen\_DGS              & 3.60 & 3.01 & 3.57 & 3.25 & 3.09 & 3.30 \\
        & Qwen\_SRS              & \textbf{3.84} & \textbf{3.51} & \textbf{3.84} &\textbf{ 3.54} & \textbf{3.35} & \textbf{3.62} \\
        
        & Qwen-Coder\_DGS        & 3.92 & 3.43 & 3.83 & 3.61 & 3.31 & 3.62 \\
        & Qwen-Coder\_SRS        & \textbf{3.94} & \textbf{3.52} & \textbf{4.10} & \textbf{3.69} & \textbf{3.53} & \textbf{3.76} \\
        \midrule
        \multirow{11}{*}{Python}
        & Reference              & 2.90 & 2.78 & 2.99 & 2.37 & 2.59 & 2.73 \\
        
        & GPT-4o\_DGS         & \textbf{4.54} & \textbf{3.89} & \textbf{4.17} &\textbf{ 3.81} & \textbf{4.01} & \textbf{4.08} \\
        & GPT-4o\_SRS         & 4.36 & 3.85 & 3.94 & 3.76 & 3.87 & 3.96 \\
        
        & DeepSeek\_DGS          & 3.80 & 4.10 & 3.54 & 3.34 & 3.45 & 3.65 \\
        & DeepSeek\_SRS          & \textbf{3.83} & \textbf{4.16} & \textbf{3.87} & \textbf{3.56} & \textbf{3.54} & \textbf{3.79} \\
        
        & DeepSeek-Coder\_DGS    & 3.85 & 3.21 & 3.83 & 3.30 & 3.43 & 3.52 \\
        & DeepSeek-Coder\_SRS    & \textbf{3.86} & \textbf{3.28} & \textbf{3.91} & \textbf{3.43} & \textbf{3.50} & \textbf{3.60} \\
        
        & Qwen\_DGS              & 3.63 & 2.97 & 3.62 & 3.26 & 3.28 & 3.35 \\
        & Qwen\_SRS              & \textbf{3.69} & \textbf{3.27} & \textbf{3.83} & \textbf{3.32} & \textbf{3.39} & \textbf{3.50} \\
        
        & Qwen-Coder\_DGS        & 3.80 & 3.34 & 3.94 & 3.42 & 3.45 & 3.59 \\
        & Qwen-Coder\_SRS        & \textbf{3.97} & \textbf{3.45} & \textbf{4.24} & \textbf{3.71} & \textbf{3.61} & \textbf{3.80} \\
        \bottomrule
        \end{tabular}
\end{table*}

\subsubsection{Experimental Results.}
Table~\ref{tab:evaluation_direct_generation_strategy_vs_summarization-and-refinement_strategy} presents the LLM-based evaluation results for reference method names and method names generated by the \textit{direct generation strategy} (DGS) and the \textit{summarization-and-refinement strategy} (SRS) across Java and Python. Overall, the LLM-generated method names receive substantially higher scores than the reference method names. For Java, the average score of the reference method names is 2.75, whereas the average scores of LLM-generated method names generally range from 3.30 to 4.02. A similar trend is observed for Python, where the reference method names obtain an average score of 2.73, while the generated method names achieve average scores between 3.35 and 4.08. These results indicate that LLM-generated method names are generally preferred by LLM evaluators over the reference names.

Among the LLM-generated method names, SRS generally achieves better performance than DGS. Except for GPT-4o, where DGS obtains the highest average scores in both Java and Python, SRS consistently improves the average scores for the other LLMs. For instance, Qwen improves from 3.30 to 3.62 in Java, and Qwen-Coder improves from 3.59 to 3.80 in Python. These results indicate that introducing an intermediate summarization step can improve the quality of generated method names in most model configurations.

\begin{figure}[htbp]
\centering
    \includegraphics[width=\columnwidth]{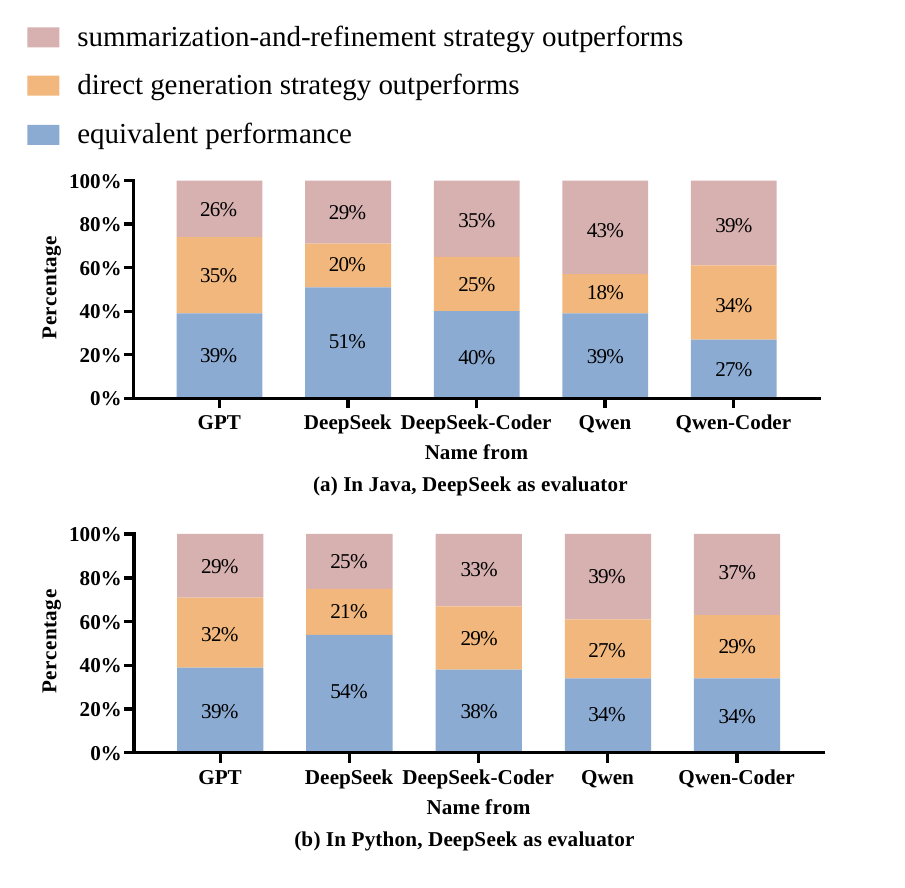}
    \caption{\protect\raggedright LLM-based preference outcomes for \textit{direct generation strategy } and \textit{summarization-and-refinement strategy} using DeepSeek as evaluator on Java and Python MNP tasks.}
    \label{fig:distributions_naming_strategies}
\end{figure}

\textbf{Distributions for Naming Strategies:} Figure~\ref{fig:distributions_naming_strategies} further illustrates the distribution of LLM-based evaluation outcomes between the \textit{direct generation strategy} and the \textit{summarization-and-refinement strategy} across Java and Python tasks, using DeepSeek as the evaluator. According to Table~\ref{tab:correlation_coefficient}, DeepSeek exhibits the strongest and most consistent correlation with human evaluation among all LLM-based evaluators, making it the most reliable proxy for human judgment. The exact proportion for each strategy is shown on the stacked bars.  Across the 200 samples, each sample falls into one of three outcome categories: the \textit{summarization-and-refinement strategy} outperforms, the \textit{direct generation strategy} outperforms, or the two strategies achieve equivalent performance. We count the occurrences of each outcome and calculate its proportion in the total sample.

As shown in Figure~\ref{fig:distributions_naming_strategies}, the \textit{summarization-and-refinement strategy} consistently outperforms the \textit{direct generation strategy} across multiple models, achieving higher preference ratios in most cases. This trend is consistent across both Java and Python, with preference for the summarization-and-refinement strategy ranging from 26\% to 43\% (Java) and 25\% to 39\% (Python), markedly higher than that of the direct generation strategy (Java: 18\%–35\%; Python: 21\%–32\%). These results provide strong evidence that the \textit{summarization-and-refinement strategy} yields consistently higher improvement proportions, validating its effectiveness in improving both the semantic accuracy and naturalness of generated method names.

\summary{These results indicate that when LLMs serve as evaluators, the \textit{summarization-and-refinement strategy} typically outperforms the \textit{direct generation strategy} across most cases. The proportion of samples improved by this strategy is consistently higher, further validating its effectiveness.}

\section{Our Methodology}
\label{sec:methodology}
Based on the above empirical studies, we further conduct an in-depth case study and identify three limitations in the naive implementation of the \textit{summarization-and-refinement strategy}: summarization semantic inaccuracy, refinement semantic misalignment, and semantic score proximity, as detailed in Section~\ref{subsec:limitations_of_naive_SRS}.
To address these limitations, we propose a novel MNP approach, namely \ours{}, detailed in Section~\ref{subsec:our_methodology}. \ours{} devises two components, i.e., MNP-oriented summarization and chain-of-thought (CoT)–enhanced refinement to address the problems of summarization semantic inaccuracy and refinement semantic misalignment, respectively.

\subsection{Limitations of Naive Implementation of SRS}
\label{subsec:limitations_of_naive_SRS}
According to the experimental results of RQ2 (Section~\ref{subsec:RQ2_comparative_effectiveness}), the analysis is based on 200 original code bodies. For each code body, five LLMs generate method names under two generation strategies, namely DGS and SRS, resulting in five DGS--SRS comparison cases per code body and 1000 comparison cases in total. Each generated method name is further evaluated by five LLM evaluators. Therefore, the following error analysis is conducted at the model-specific comparison-case level rather than at the unique code-body level. To mitigate single-model bias, we adopt a voting-like mechanism and further consider the average evaluation score across the five LLM evaluators. Specifically, for each DGS--SRS comparison case, we compare the SRS-generated method name with the corresponding DGS-generated method name across the five LLM evaluators. If the SRS-generated method name receives higher scores from at least three of the five evaluators, we consider SRS to outperform DGS for that case. The average score across the five evaluators is also used as a complementary indicator to assess the overall difference in quality between the two strategies.

We then examine the cases where SRS does not clearly outperform DGS and analyze their corresponding intermediate summaries and generated method names. Through this analysis, we identify 423 limitation cases in total, which are categorized into three major types: (1) summarization semantic inaccuracy; (2) refinement semantic misalignment, including semantic object misalignment and semantic verb misalignment; and (3) semantic score proximity. Specifically, summarization semantic inaccuracy includes 64 cases where the generated summary deviates from the intended functionality of the method. Refinement semantic misalignment includes 265 cases in total, consisting of 99 cases of semantic object misalignment and 166 cases of semantic verb misalignment. Semantic score proximity includes 94 cases where the compared SRS- and DGS-generated method names receive identical or highly similar evaluation scores, making it difficult to determine which strategy performs better for those cases.

\noindent\textbf{(1) Summarization Semantic Inaccuracy:}
The summary generally captures the code's functionality, but it contains minor factual errors and inaccuracies in the details. Such discrepancies mislead the subsequent method name extraction process, resulting in suboptimal naming decisions.
\begin{figure}[htbp]
    \centering
    \includegraphics[width=\columnwidth]{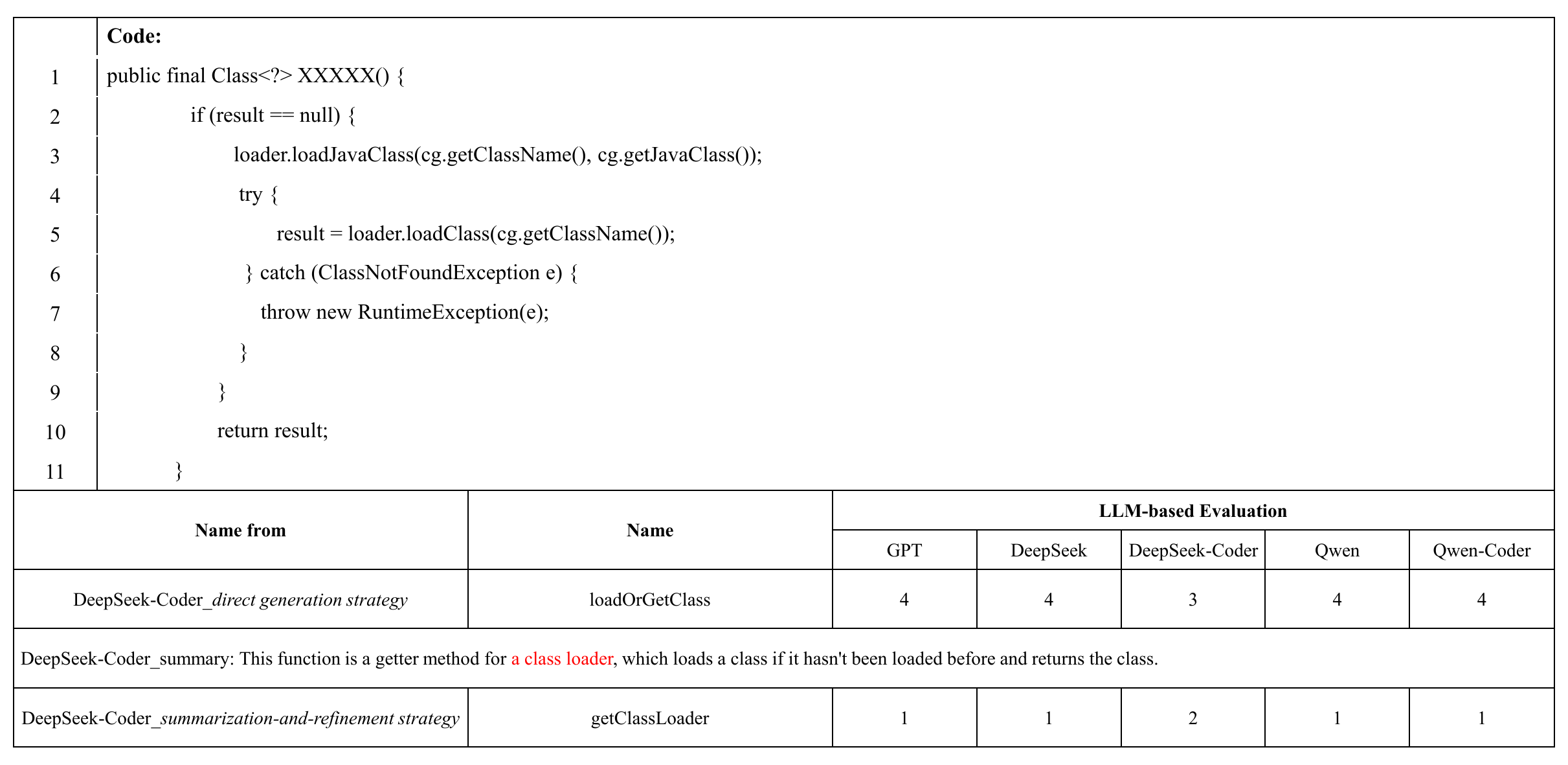}
    \caption{Example of a summarization semantic inaccuracy}
\label{fig:example_summarization_intent_divergence}
\end{figure}

In Figure~\ref{fig:example_summarization_intent_divergence}, the method conditionally loads and returns a Java class: if the class is not loaded (result == null), it invokes the loader to load the class and subsequently returns the loaded Class object. The corresponding DeepSeek-Coder summary erroneously characterizes the function as "a getter method for a class loader," inaccurately suggesting that the return value is a ClassLoader rather than a Class. This semantic inaccuracy propagates to the automated naming process, as evidenced by the evaluation scores. The name "loadOrGetClass", produced under the \textit{direct generation strategy}, partially captures the function's behavior, yielding relatively high scores across GPT, DeepSeek, and Qwen evaluators (4, 4, and 4, respectively). By contrast, the name "getClassLoader," generated under the \textit{summarization-and-refinement strategy}, misrepresents the return type and consequently receives substantially lower scores (1–2). These results demonstrate that misleading summaries, particularly those containing factual inaccuracies, omitted details, or ambiguous descriptions, systematically induce method names that are semantically imprecise and incomplete, thereby undermining their practical utility. This case highlights the critical role of summary accuracy in automated method naming: even minor misrepresentations of functional behavior can lead to incorrect or suboptimal method names, underscoring the necessity of precise semantic descriptions in downstream name generation.

\noindent\textbf{(2) Refinement Semantic Misalignment:}
During the refinement phase, the model exhibits two types of semantic mismatches between the method name and its intended functionality: 1) semantic verb misalignment, where the model employs an inappropriate verb that fails to reflect the correct action; and 2) semantic object misalignment, where the model extracts an inaccurate object that does not correctly represent the target entity or data being modified.

\begin{figure}[htbp]
    \centering
    \includegraphics[width=\columnwidth]{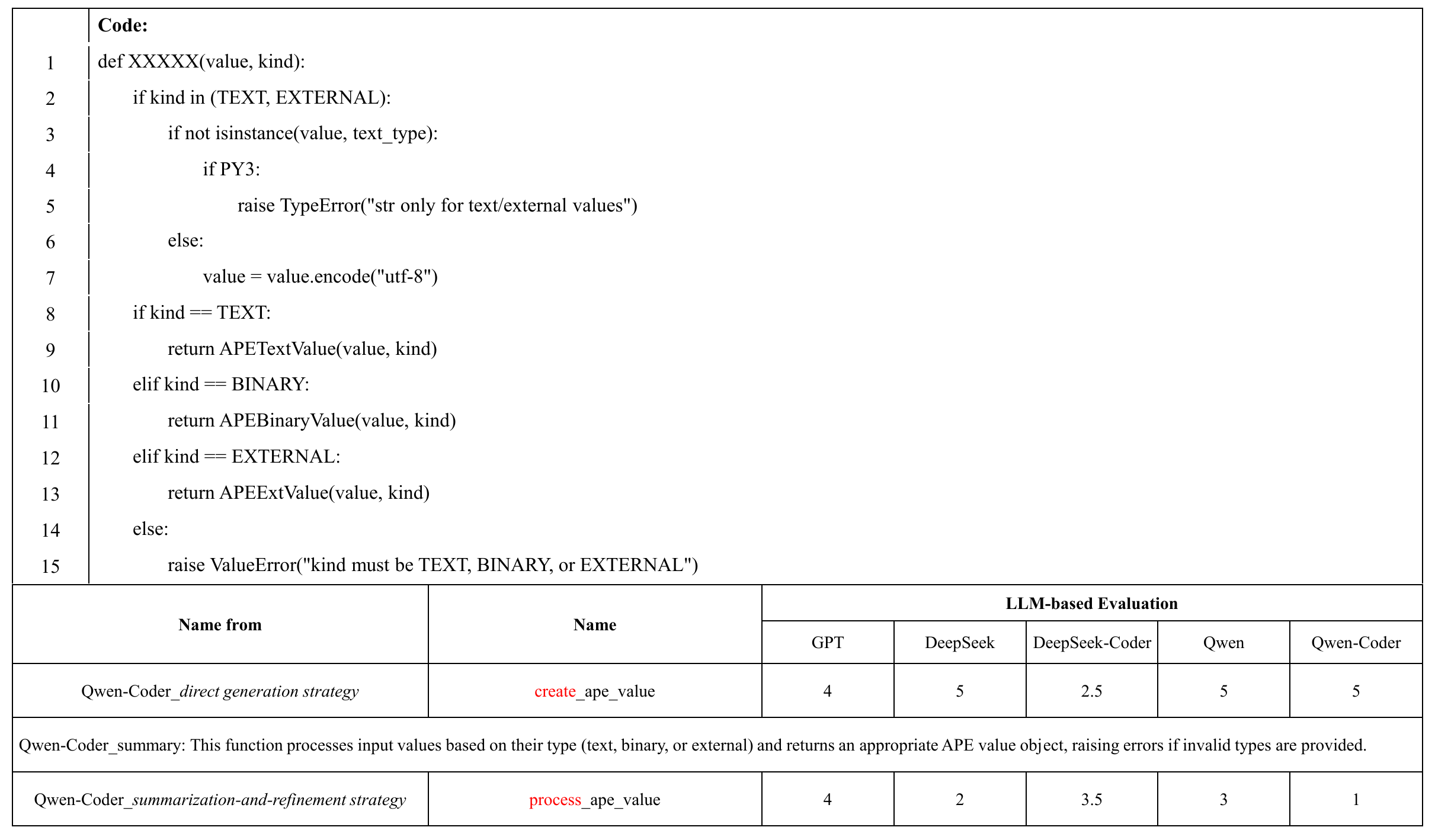}
    \caption{Example of a Semantic Verb Misalignment}
\label{fig:example_name_verb}
\end{figure}

\textbf{1) Semantic Verb Misalignment: }In Figure~\ref{fig:example_name_verb}, the summary correctly captures the function's actions, stating that it "processes ... and returns ..." an APE value object. The combination of "processes" and "returns" implicitly suggests a constructive semantics: the function creates a new object. However, during the refinement phase, the model fails to infer the implicit meaning and directly adopts the verb "process" from the summary, yielding the method name "process\_ape\_value." This represents a semantic verb misalignment: the model does not refine the coarse-grained verb "process" into the more precise verb "create" that better reflects the function's true behavior. By contrast, the DGS produces "create\_ape\_value," achieving higher evaluation scores (DeepSeek: 5, Qwen: 5, Qwen-Coder: 5) compared to the SRS name (DeepSeek: 2, Qwen: 3, Qwen-Coder: 1). This case demonstrates that even when a summary contains sufficient semantic information, the naive refinement phase may fail to extract and generalize the most appropriate verb, leading to suboptimal naming outcomes.

\begin{figure}[htbp]
    \centering
    \includegraphics[width=\columnwidth]{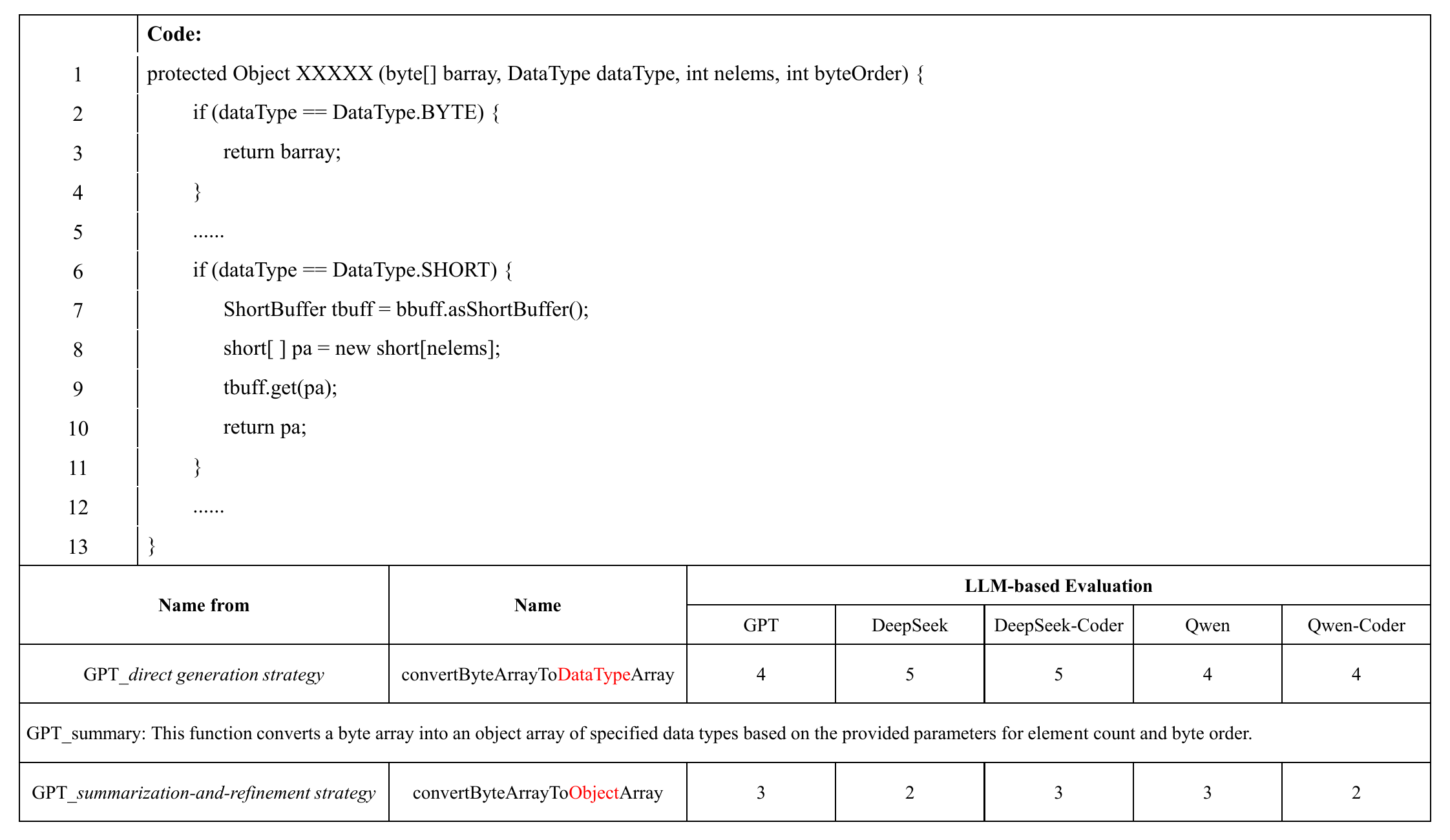}
    \caption{Example of a Semantic Object Misalignment}
\label{fig:example_name_object}
\end{figure}
\textbf{2) Semantic Object Misalignment: }In Figure~\ref{fig:example_name_object}, the summary correctly states that the function converts a byte array into an object array of specified data types. However, during the refinement phase, the model exhibits semantic object misalignment: although the summary explicitly states that the function returns an object array of specified data types, the model fails to further refine this information. It directly extracts "ObjectArray" as the target object without specializing it to "DataTypeArray," producing the method name "convertByteArrayToObjectArray." In other words, the model does not refine the nuanced "specified data types" into a more precise "DataTypeArray" but instead retains a generic "ObjectArray" description. The DGS produces "convertByteArrayToDataTypeArray," achieving consistently higher evaluation scores across all five models compared to the SRS name. This case demonstrates that semantic object misalignment occurs when the model extracts an imprecise or overly general object from the summary, thereby failing to leverage the specificity provided in the summary, leading to method names that do not accurately reflect the function's actual return-type semantics.

\noindent\textbf{(3) Semantic Score Proximity: }
The generated names use synonyms or semantically equivalent words that closely approximate the intended meaning, leading LLM evaluators to assign similar scores to both strategies and making it difficult to distinguish their relative quality.
\begin{figure}[htbp]
    \centering
    \includegraphics[width=\columnwidth]{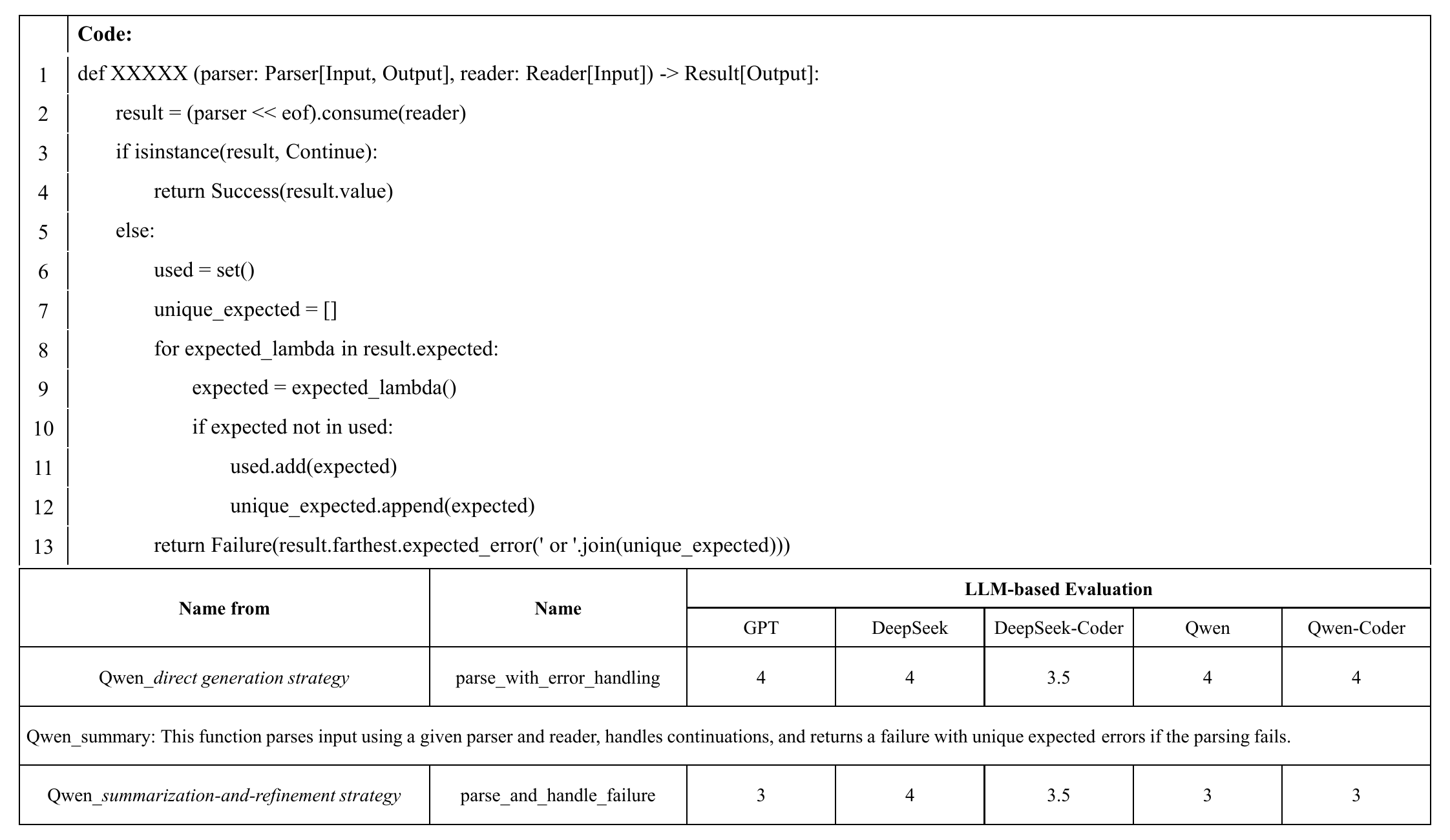}
    \caption{Example of a Semantic Similarity Issue}
\label{fig:example_same}
\end{figure}

In Figure~\ref{fig:example_same}, the summary correctly captures the function's behavior, stating that it "parses input using a given parser and reader, handles continuations, and returns a failure with unique expected errors if parsing fails." During the refinement phase, the SRS generates the name "parse\_and\_handle\_failure," while the DGS produces "parse\_with\_error\_handling." These two names are semantically close: both convey the idea of parsing and managing errors or failures. The similarity in meaning is reflected in the evaluation scores, where the SRS name performs comparably to the DGS, with no substantial gaps across most models. This case illustrates semantic score proximity: the SRS name, while not identical to the DGS name, carries a similar semantic intent and achieves acceptable quality. This phenomenon is less critical than summarization semantic inaccuracy and refinement semantic misalignment, as it does not fundamentally misrepresent the function's behavior.

\subsection{\ours{}}
\label{subsec:our_methodology}
To address the two predominant limitations of the naive SRS implementation, including summarization semantic inaccuracy and refinement semantic misalignment, we propose an enhanced approach comprising two key components: MNP-Oriented Summarization and CoT-Enhanced Refinement.

\noindent\textbf{(1) MNP-Oriented Summarization:}
During the code summarization stage, we introduce an MNP-oriented summarization mechanism to generate summaries that are more suitable for downstream method name prediction. The prompt is designed to guide the LLM to focus on the method's core value to its caller, including the main action, the primary object of operation, and the key result or effect. It also encourages the model to preserve key domain terms while avoiding low-level implementation details, overly generic verbs, and vague descriptions. 
The prompt template is as follows:
\textit{''
Given the function code, write ONE concise comment sentence for method-name generation that describes what the function does for its caller, stating its main action, the main object it operates on, and the main result or effect, while keeping key domain terms. Avoid low-level implementation details, avoid generic verbs such as ``do'' or ``execute'', and avoid descriptions that simply start with ``return(s)''. Do not use ``create'' unless it truly creates a new externally visible entity. Output only a single-line comment starting with the appropriate comment marker.''}

\noindent\textbf{(2) CoT-Enhanced Refinement:}
During the method name generation stage, we incorporate CoT prompting to refine the intermediate summary into a more accurate method name. The prompt first assigns the LLM the role of an experienced software developer and provides two inputs: the programming language and the functional summary generated in the previous stage. It then guides the model through a six-step reasoning process before producing the final method name. Among these steps, Steps 1--3 constitute the core semantic reasoning process, focusing on the method's core purpose, key action, and target entity, respectively. Steps 4--6 further guide the model to apply language-specific naming conventions, synthesize the final method name, and output it in the required format.

\textbf{Step 1: Understand the Core Purpose.}
This step requires the model to identify the high-level goal and responsibility of the method based on the summary. It helps establish the overall semantic intent before selecting specific naming tokens.

\textbf{Step 2: Identify the Key Action.}
This step guides the model to determine the primary operation performed by the method, such as validation, transformation, calculation, filtering, or retrieval. Making the action explicit helps reduce the risk of selecting an inaccurate or overly generic verb in the generated method name.

\textbf{Step 3: Determine the Target Entity.}
This step requires the model to identify the main object, data, or conceptual entity involved in the operation. It helps ensure that the generated name reflects both the action and the object being acted upon.

By explicitly reasoning about the purpose, action, and target entity before applying naming conventions and producing the final output, the CoT-enhanced refinement mechanism helps mitigate semantic verb misalignment and semantic object misalignment in the refinement stage.

\section{Results and Evaluation}
\label{sec:results_and_evaluation}

\subsection{RQ3: What is the overall effectiveness of the \ours{} for MNP?} 

\subsubsection{Experimental Setup}

In this RQ, we evaluate \ours{} against SRS using five LLM evaluators: GPT, DeepSeek, DeepSeek-Coder, Qwen, and Qwen-Coder. Following the preceding empirical study, we conduct the evaluation on the 423 limitation cases identified from the naive SRS analysis, which cover summarization semantic inaccuracy, refinement semantic misalignment, and semantic score proximity. Although \ours{} mainly targets the first two error-oriented limitations, all three types are retained to assess their overall effect on naive SRS. The same reference method names, evaluation settings, and temperature setting of 0 are used.

\begin{table*}[!t]
    \centering
    \scriptsize
    \tabcolsep=12pt
   \caption{Comparison between the \textit{summarization-and-refinement strategy} (SRS) and SMNP using LLMs as evaluators}
    \label{tab:evaluation_srs_vs_smnp}
    \begin{tabular}{llcccccc}
        \toprule
        \multirow{2}{*}{\textbf{Language}} 
        & \multirow{2}{*}{\textbf{Name Source}} 
        & \multicolumn{5}{c}{\textbf{LLM-based Evaluation}} \\
        \cmidrule(lr){3-8}
        & & GPT-4o & DeepSeek & DeepSeek-Coder & Qwen & Qwen-Coder & Average \\
        \midrule
        \multirow{11}{*}{Java}
        & Reference             & 2.94 & 2.72 & 3.00 & 2.53 & 2.58 & 2.76 \\

        & GPT-4o\_SRS            & 4.27 & 3.80 & 3.58 & 3.52 & 3.32 & 3.70 \\
       & GPT-4o\_SMNP            & \textbf{4.28} & \textbf{3.97} & \textbf{3.87} & \textbf{3.78} & \textbf{3.62} & \textbf{3.90} \\

        & DeepSeek\_SRS          & 3.30 & 3.30 & 3.10 & 2.97 & 2.80 & 3.09\\
        & DeepSeek\_SMNP         & \textbf{3.83} & \textbf{3.60} & \textbf{3.97} & \textbf{3.60} & \textbf{3.67} & \textbf{3.73} \\

        & DeepSeek-Coder\_SRS    & 3.18 & 2.45 & 3.47 & 2.82 & 2.92 & 2.97 \\
        & DeepSeek-Coder\_SMNP   & \textbf{3.74} & \textbf{3.37} & \textbf{3.79} & \textbf{3.47} & \textbf{3.47} & \textbf{3.57} \\

        & Qwen\_SRS              & 3.30 & 3.10 & 3.40 & 3.03 & 3.00 & 3.17\\
       & Qwen\_SMNP              & \textbf{3.70} & \textbf{3.30} & \textbf{3.73} & \textbf{3.53} & \textbf{3.27} & \textbf{3.51} \\

        & Qwen-Coder\_SRS        & 4.00 & 3.69 & 4.02 & 3.82 & 4.04 & 3.91 \\
       & Qwen-Coder\_SMNP        & \textbf{4.02} & \textbf{3.78} & \textbf{4.18} & \textbf{3.96} & \textbf{4.14} & \textbf{4.02} \\

        \midrule
        \multirow{11}{*}{Python}
        & Reference              & 2.84 & 2.87 & 3.00 & 2.42 & 2.68 &2.76\\    

        & GPT-4o\_SRS            & 4.20 & 3.58 & 3.69 & 3.47 & 3.54 & 3.70\\
        & GPT-4o\_SMNP           & \textbf{4.42} & \textbf{3.73} & \textbf{3.93} & \textbf{3.56} & \textbf{3.66} & \textbf{3.86} \\

        & DeepSeek\_SRS          & 3.57 & 3.64 & 3.61 & 3.18 & 2.96 &3.39\\
        & DeepSeek\_SMNP         & \textbf{3.71} & \textbf{4.11} & \textbf{3.86} & \textbf{3.43} & \textbf{3.43} & \textbf{3.71} \\

        & DeepSeek-Coder\_SRS    & 3.23 & 2.79 & 3.58 & 2.93 & 2.91 & 3.09\\
        & DeepSeek-Coder\_SMNP   & \textbf{3.65} & \textbf{3.21} & \textbf{3.86} & \textbf{3.40} & \textbf{3.53} & \textbf{3.53} \\      

        & Qwen\_SRS              & 3.14 & 3.00 & 3.42 & 2.98 & 3.07 &3.12\\
        & Qwen\_SMNP             & \textbf{3.51} & \textbf{3.44} & \textbf{4.35} & \textbf{3.67} & \textbf{3.44} & \textbf{3.68} \\       

        & Qwen-Coder\_SRS        & 3.21 & 2.85 & 3.82 & 2.95 & 2.87 &3.14\\
        & Qwen-Coder\_SMNP      & \textbf{3.87} & \textbf{3.51} & \textbf{4.28} & \textbf{3.72} & \textbf{3.59} & \textbf{3.79} \\
        \bottomrule
        \end{tabular}

\end{table*}

\subsubsection{Experimental Results.}

We first examine the limitation-resolution results. Since semantic score proximity mainly reflects indistinguishable evaluation outcomes rather than clear semantic errors, the 94 cases in this category are excluded from the resolution-rate analysis. For this category, \ours{} improves the average score by only 0.04, indicating that these cases do not involve substantial quality degradation in generated method names.
For the two error-oriented categories, \ours{} shows clear improvements. It resolves 50 out of 64 cases of summarization semantic inaccuracy, achieving a resolution rate of 78\%. For refinement semantic misalignment, \ours{} resolves 75 out of 99 semantic object misalignment cases and 116 out of 166 semantic verb misalignment cases, corresponding to resolution rates of 76\% and 70\%, respectively. The average scores also increase by 0.55, 0.37, and 0.50 for summarization semantic inaccuracy, semantic object misalignment, and semantic verb misalignment. 

Table~\ref{tab:evaluation_srs_vs_smnp} further reports the LLM-based evaluation scores. Both SRS and \ours{} outperform the reference method names, while \ours{} achieves higher average scores than SRS in every paired comparison. This shows that \ours{} consistently improves upon naive SRS across all evaluated generation models and both programming languages. The gains are more pronounced in Java, suggesting that MNP-oriented summarization and CoT-enhanced refinement are effective in improving the semantic quality of generated method names. GPT-4o maintains relatively high scores under both SRS and \ours{}, indicating its strong baseline capability for method name generation. Nevertheless, \ours{} still improves upon SRS, showing that the proposed enhancements remain beneficial even for stronger LLMs.

\summary{The results show that both SRS and \ours{} outperform the reference method names, while \ours{} further improves upon SRS in every paired comparison. The improvements are more pronounced in Java, and GPT-4o maintains strong performance in both programming languages.}

\subsection{RQ4: What is the impact of removing each optimization component on the performance of the \ours{}?} 

\subsubsection{Experimental Setup}
This RQ investigates the impact of each optimization component in the proposed \ours{} framework through an ablation study. The experimental environment and configuration are kept consistent with those defined in RQ3 to ensure a fair and comparable assessment. Based on the results from Table~\ref{tab:evaluation_srs_vs_smnp}, which shows that the \ours{} approach achieves its highest performance with GPT-4o, we select GPT-4o as the model to generate method names for this ablation study.

\begin{table*}[t]
\centering
\tiny
\caption{Ablation Study of the SMNP Framework on Java and Python (Method Names Generated by GPT-4o)}
\label{tab:ablation_study}
\resizebox{\textwidth}{!}{
\begin{tabular}{clcccccc}
\toprule
\multirow{2}{*}{\textbf{Language}} 
& \multirow{2}{*}{\textbf{Variant}} 
& \multicolumn{6}{c}{\textbf{LLM-based Evaluation}} \\
\cmidrule(lr){3-8}
& & GPT-4o & DeepSeek & DeepSeek-Coder & Qwen & Qwen-Coder & AVG \\
\midrule

\multirow{6}{*}{\textbf{Java}}
& \textit{Summarization-and-refinement Strategy}
& 4.37 & 3.89 & 3.90 & 3.76 & 3.78 & 3.94 \\

& w/o Task-Oriented summary
& 4.34 & 3.90 & 3.91 & 3.76 & 3.80 & 3.94 \\

& w/o CoT Step 1: Understand the Core Purpose
& 4.29 & 3.86 & 3.91 & 3.78 & 3.73 & 3.91 \\

& w/o CoT Step 2: Identify Key Actions
& 4.32 & 3.78 & 3.89 & 3.63 & 3.76 & 3.88 \\

& w/o CoT Step 3: Determine the Target Entity
& 4.36 & 3.86 & 3.90 & 3.71 & 3.78 & 3.92 \\

& \textbf{SMNP (ours)}
& \textbf{4.38} & \textbf{4.17} & \textbf{3.92} & \textbf{4.07} & \textbf{4.20} & \textbf{4.15} \\
\midrule

\multirow{6}{*}{\textbf{Python}}
& \textit{Summarization-and-refinement Strategy}
& 4.36 & 3.85 & 3.94 & 3.76 & 3.87 & 3.96 \\

& w/o Task-Oriented summary
& 4.28 & 3.86 & 3.97 & 3.80 & 3.81 & 3.94 \\

& w/o CoT Step 1: Understand the Core Purpose
& 4.23 & 3.82 & 3.92 & 3.63 & 3.75 & 3.87 \\

& w/o CoT Step 2: Identify Key Actions
& 4.27 & 3.86 & 3.97 & 3.75 & 3.81 & 3.93 \\

& w/o CoT Step 3: Determine the Target Entity
& 4.30 & 3.87 & 3.88 & 3.69 & 3.78 & 3.90 \\

& \textbf{SMNP (ours)}
& \textbf{4.47} & \textbf{4.15} & \textbf{4.03} & \textbf{4.07} & \textbf{4.22} & \textbf{4.19} \\
\bottomrule
\end{tabular}
}
\end{table*}

\subsubsection{Experimental Results}
Table~\ref{tab:ablation_study} compares variants with each component removed in turn, allowing the relative importance of each part to be assessed.:

\textbf{(1) Step 2: Identify Key Actions is the most critical}. Ablating this step leads to the most pronounced decline in average score across both languages (Java falls to 3.88, Python to 3.93). This result strongly supports our earlier diagnosis that Semantic Verb Misalignment is a central obstacle to generating high‑quality method names, and that explicitly isolating the core action verb is the most effective countermeasure.

\textbf{(2) Step 1: Understand the Core Purpose and Step 3: Determine the Target Entity play significant and complementary roles}. Removal of either step also reduces the average score. Step 1 anchors the naming process to the method’s overarching intent, while Step 3 ensures the primary operand or object is accurately reflected. Together, they systematically address Semantic Object Misalignment, enabling a complete and precise semantic abstraction of the method’s operation.

\summary{This structured ablation confirms that the full \ours{} framework, integrating an intermediate summary and explicit CoT reasoning, is essential for generating semantically precise names, and that the explicit identification of the key action in Step 2 is the single most impactful component.}

\section{Threats to Validity}
\label{sec:Threats to validity}

Our study may face several threats to validity, which we have attempted to mitigate.

\textbf{Threats to Internal Validity.}
The main threats to internal validity involve potential implementation errors in the evaluation pipeline and the processing of LLM outputs. To reduce this risk, we rely on publicly available implementations for traditional automated metrics and use official scripts or well-maintained APIs for the LLMs involved in this study. Another threat is the extraction of final method names or code summaries from LLMs' raw textual responses, which may contain extra explanatory text. To ensure consistent and accurate extraction, we design and apply strict heuristic rules, including pattern matching based on naming conventions and content extraction based on fixed output formats~\cite{2025-Summarization-LLM}.

\textbf{Threats to External Validity.}
A primary threat to external validity arises from the inherent randomness of LLMs, which may generate different outputs for the same input across multiple runs. To mitigate this threat, we set the temperature parameter to 0 during both generation and LLM-based evaluation to reduce output randomness and improve scoring consistency. We also evaluate our approach on two programming languages, Java and Python, to improve the robustness of our conclusions across different contexts. However, our findings are based on a specific dataset of 200 samples, and their applicability to other domains, programming languages, or substantially different codebases requires further investigation.

\section{Conclusion}
\label{sec:conclusion}
In this paper, we propose \ours{}, a method that leverages code summarization as a semantic bridge for method name generation, decomposing code understanding into a "code-summary-name" chain-reasoning task. Through systematic experimentation, this study draws the following key conclusions: 1) LLM-based evaluation methods exhibit stronger and more stable correlation with human assessment than traditional automated metrics, confirming their suitability for evaluating generated method names. 2)The \textit{summarization-and-refinement strategy} (SRS), which simulates the human "understanding-before-naming" process, produces higher quality method names than the \textit{direct generation strategy} (DGS). 3)By addressing the identified limitations of naive SRS, namely summarization semantic inaccuracy and refinement semantic misalignment, \ours{} effectively improves method name quality, with the improvement being consistently more pronounced than that of SRS.

\section{Acknowledgements}
\thanks{The authors would like to thank the anonymous reviewers for their insightful comments. 
This work is supported in part by the National Natural Science Foundation of China under Grant 62272203 and in part by the Joint Funds of the National Natural Science Foundation of China under Grant U24A20238.}

\bibliographystyle{IEEEtran}
\bibliography{reference}

\end{document}